\documentclass[aps,prl,superscriptaddress,amsmath,groupedaddress,twocolumn]{revtex4}

\usepackage{color,hangcaption,hhline,psfrag,rotating,amssymb}
\usepackage[hang,nooneline]{subfigure}
\usepackage{dcolumn}

\begin{document}
\title{High Precision determination of the $\pi$, $K$, $D$ and $D_s$ decay constants from lattice QCD}

\author{E. Follana}
\author{C. T. H. Davies}
\email[]{c.davies@physics.gla.ac.uk}
\affiliation{Department of Physics and Astronomy, University of Glasgow, Glasgow, G12 8QQ, UK}
\author{G. P. Lepage}
\affiliation{Laboratory of Elementary-Particle Physics, Cornell University, Ithaca, New York 14853, USA}
\author{J. Shigemitsu}
\affiliation{Department of Physics, The Ohio State University, Columbus, Ohio 43210, USA}

\collaboration{HPQCD and UKQCD collaborations}
\noaffiliation

\date{\today}

\begin{abstract}
We determine $D$ and $D_s$ decay constants from 
lattice QCD with 2\% errors, 4 times better than experiment 
and previous theory: $f_{D_s}$ = 241(3) MeV, $f_D$ = 207(4) MeV and 
$f_{D_s}/f_D$ = 1.164(11). 
We also obtain $f_K/f_{\pi}$ = 1.189(7) and
$(f_{D_s}/f_D)/(f_K/f_{\pi})$ = 0.979(11). Combining 
with experiment gives $V_{us}$=0.2262(14) 
and 
$V_{cs}/V_{cd}$ of 4.43(41). 
We use a highly improved quark discretisation 
on MILC gluon fields that
include realistic sea quarks, 
fixing the $u/d, s$ and $c$ masses from the $\pi$, $K$, and $\eta_c$ meson 
masses. This allows a stringent test against experiment 
for $D$ and $D_s$ masses for the first time (to within 7 MeV). 
\end{abstract}


\maketitle

The annihilation to a $W$ boson of the $D_s$, $D_d$, $\pi$ or $K$ meson is a `gold-plated' process 
with leptonic width (for meson $P$ of quark content $a\overline{b}$) given, up to a calculated 
electromagnetic correction factor~\cite{marciano, pdg06}, by:
\begin{equation}
\Gamma(P \rightarrow l \nu_l (\gamma)) = \frac{G_F^2 |V_{ab}|^2}{8\pi}f_{P}^2m_l^2m_{P}\left( 1-\frac{m_l^2}{m_{P}^2}\right)^2.
\label{eq:gamma}
\end{equation}
$V_{ab}$ is from the Cabibbo-Kobayashi-Maskawa (CKM) 
matrix and
$f_{P}$, the decay constant, parameterizes the amplitude for $W$
annihilation. 
If $V_{ab}$ is known from elsewhere
an 
experimental value for $\Gamma$ gives $f_{P}$, 
to be compared to theory. 
If not, an accurate theoretical value for $f_{P}$, 
combined with experiment, can yield a value 
for $V_{ab}$. 

$f_{P}$ is defined from
$\langle 0 | \overline{a} \gamma_{\mu} \gamma_5 b | P(p) \rangle \equiv f_{P} p_{\mu}$
calculable in lattice QCD to handle quark confinement, and with QED effects omitted.
The experimental leptonic decay rates for $K$ and $\pi$ are known very 
accurately and 
$D$ and $D_s$ less so, but with expected errors shortly of a 
few percent. Accurate predictions from lattice 
QCD can be made now, ahead of these results 
and comparison will then be a severe test of lattice QCD
(and QCD itself). This has impact on the confidence we 
have in similar matrix elements 
being calculated in lattice QCD for $B$ mesons that
provide key unitarity triangle constraints. 

A major error in lattice QCD 
until recently was missing
sea quarks from the 
gluon field configurations on which calculations
were done, because of numerical 
expense. 
This has now been overcome. 
The MILC collaboration~\cite{milc2} 
has made ensembles at several different values of the lattice 
spacing, $a$, that include sea $u$ and $d$ (taken to have the same 
mass) and $s$ quarks with the $u/d$ quark mass taking a range of 
values down to $m_s/10$. The sea quarks are implemented in the improved 
staggered (asqtad) formalism by use of the fourth root of the quark 
determinant. This procedure, although deemed `ugly', appears to be 
a valid discretisation of QCD~\cite{sharpe}. 

For $f_{D_q}$ a large error can arise from the inaccuracy of 
discretisations of QCD 
for $c$ quarks. 
Discretisation errors are set by powers of the mass in lattice units, $m_ca$, 
and this is not negligible at typical values of $a$.
However, $m_ca$ is not so large that it can easily be 
removed from the problem using nonrelativistic methods
as is done for $b$ quarks, for example~\cite{nrqcd}. 
The key then to obtaining small errors for 
$c$ quarks is a highly improved relativistic action on reasonably
fine lattices (where $m_ca \approx 1/2$). 

\begin{table}
\begin{tabular}{l|l|l}
Lattice/sea & valence & $r_1/a$ \\
$u_0am_l$, $u_0am_s$ & $am_l$, \,\,\, $am_s$, \,\, $am_c$, $1+\epsilon$ & \\
\hline
$16^3 \times 48$ & & \\
0.0194, 0.0484 & 0.0264, 0.066, 0.85, 0.66 & 2.129(11) \\
0.0097, 0.0484 & 0.0132, 0.066, 0.85, 0.66 & 2.133(11) \\
$20^3 \times 64$ & & \\
0.02, 0.05 & 0.0278, 0.0525, 0.648, 0.79 & 2.650(8) \\
0.01, 0.05 & 0.01365, 0.0546, 0.66, 0.79 & 2.610(12) \\
$24^3 \times 64$ & & \\
0.005, 0.05 & 0.0067, 0.0537, 0.65, 0.79 &  2.632(13) \\
$28^3 \times 96$ & & \\
0.0124, 0.031 & 0.01635, 0.03635, 0.427, 0.885 & 3.711(13) \\
0.0062, 0.031 & 0.00705, 0.0366, 0.43, 0.885 & 3.684(12) \\
\end{tabular}
\caption{MILC configurations and mass parameters used for this analysis. The $16^3 \times 48$ lattices
are `very coarse', the $20^3 \times 64$ and $24^3 \times 64$, `coarse' and the 
$28^3 \times 96$, `fine'. The sea 
asqtad quark masses ($l = u/d$) are given in the MILC convention with $u_0$ the plaquette 
tadpole parameter. Note that the sea $s$ quark masses on fine and coarse lattices 
are above the subsequently determined physical value~\cite{milc3}. 
We make a 
small correction (with 50\% uncertainty) 
to our results for $f_{\pi, K, D, D_s}$ to allow for this, 
based on our studies of their sea quark mass dependence and MILC results 
in~\cite{milcsea}. It has 
negligible effect on our final numbers and errors. 
The lattice spacing values in units of $r_1$ after `smoothing'
are in the rightmost column~\cite{milc2, priv}. The central column gives the 
HISQ valence $u/d$, $s$ and $c$ masses along with the coefficient of the Naik term, 
$1+\epsilon$, used for $c$ quarks~\cite{taste}. }
\label{tab:params}
\end{table}

The FNAL and MILC collaborations 
previously obtained a prediction for $f_D$ of 201(17) MeV and 
for $f_{D_s}$ 
of 249(16) MeV~\cite{fnal-fds} using the `clover' action for 
$c$ quarks.
The 6\%-8\% error comes largely from discretization errors in the clover action. 
Our action is improved to a higher order in the lattice spacing 
and this means that our results are both more accurate for the 
$D$ and $D_s$ and more accurate for charmonium, allowing
additional predictive power. 
In addition, our formalism 
has a partially conserved current so
we do not have to renormalise the lattice $f_{D_q}$ 
to give a result for the continuous real world of experiment. 
We can then reduce the error on $f_{D_q}$ to 
2\% and the ratio of decay constants even further. 

We use the Highly Improved Staggered Quark (HISQ) action, 
developed~\cite{taste} from the asqtad 
action by reducing by a factor of 3 the `taste-changing' 
discretisation errors. 
Other discretisation errors are also small since, in common with 
asqtad, HISQ includes a `Naik' term to cancel standard tree-level $a^2$ errors in the 
discretisation of the Dirac derivative. 
For $c$ quarks the largest remaining discretisation error comes from 
radiative and tree-level corrections to the Naik term and 
we remove these by tuning the coefficient of the Naik term to obtain a 
`speed of light' of 1 in the meson dispersion relation. 
The hadron mass is then given accurately by its energy at zero momentum,
unlike the case for the clover action.   

In~\cite{taste} we tested the HISQ action extensively 
in the charmonium sector, fixing $m_c$
so that the mass of the `goldstone'
$\eta_c$ meson agreed with experiment. 
We showed that remaining discretisation 
errors are very small, being suppressed
by powers of the velocity of the $c$ quark beyond 
the formal expectation of $\alpha_s(m_ca)^2$ and $(m_ca)^4$. 
The charm quark masses 
and Naik coefficients used for different `very coarse', `coarse' and `fine' MILC ensembles 
are given in Table~\ref{tab:params}. 
For the $s$ and $u/d$ valence quarks we also use the 
HISQ action with masses in Table~\ref{tab:params}. 

We calculate local two-point 
goldstone pseudoscalar correlators at zero momentum from a precessing random 
wall source~\cite{milc3} 
and fit the average to:
\begin{eqnarray}
C(t) = \sum_{i,ip} a_ie^{-M_it}+(-1)^ta_{ip}e^{-M_{ip}t} + (t \rightarrow T-t).
\label{eq:fit}
\end{eqnarray}
$T$ is the time length of the lattice and $t$ runs from 
0 to $T$. 
$i$ denotes `ordinary' exponential terms 
and $ip$, `oscillating'
terms from opposite parity states.
Oscillating terms are significant for $D_q$ states
because $m_c-m_q$ is relatively large, but not for $\pi$ or $K$. 
We use a number of exponentials, $i$ and $ip$, in the range 2-6 
and loosely constrain higher 
order exponentials by the use of Bayesian priors~\cite{bayesfits}. 
Constraining the $D$ and $D_s$ radial excitation 
energies to be similar improves the errors on the $D$. 
In lattice units, $M_{P} = M_0$ and
$f_{P}$ is related to $a_0$ through the partially conserved 
axial current relation, 
which gives 
$f_{P} = (m_a+m_b)\sqrt{2a_0/M_0^3}$.
The resulting fitting error for all states is 
less than 0.1\% on $M_0$
in lattice units and less than 0.5\% on decay constants. 
Full details will be given in a longer paper~\cite{usinprep}. 

To convert the results to physical units 
we use the scale
determined by the MILC collaboration (Table~\ref{tab:params},~\cite{milc2}) 
in terms of the heavy 
quark potential parameter, $r_1$. 
$r_1/a$
is determined with an error of less than 0.5\% and allows 
results to be tracked accurately as a function of sea $u/d$ quark 
mass and lattice spacing. At the end, however, there is a 
larger uncertainty from 
the physical value of $r_1$. This is obtained 
from the $\Upsilon$ spectrum using the nonrelativistic QCD action 
for $b$ quarks
on the same MILC ensembles~\cite{agray}, giving $r_1$ = 0.321(5) fm, 
$r_1^{-1}=0.615(10)$ GeV. 

\begin{figure}[h]
\includegraphics[width=8.5cm,height=7.0cm]{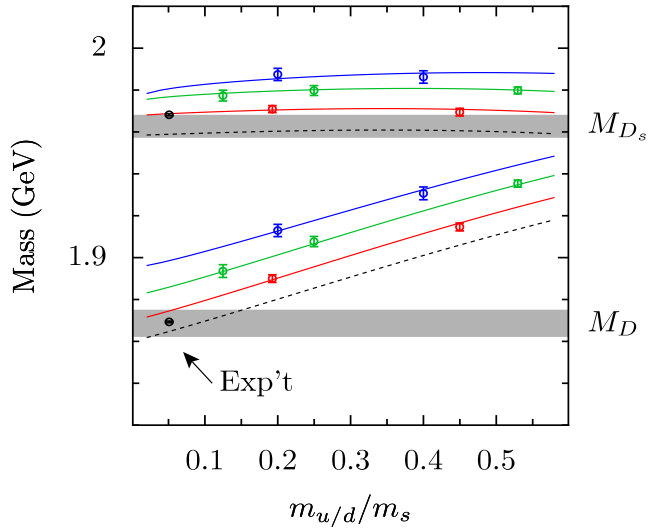}
\caption{Masses of the $D^+$ and $D_s$ meson as a function 
of the $u/d$ mass in units of the $s$ mass at three values of 
the lattice spacing. The very coarse results are the top ones in each set, then coarse, then fine.  
The lines give the 
simultaneous chiral fits and the dashed line the continuum extrapolation 
as described in the text. Our final error bars, including the 
overall scale uncertainty, are given by the shaded bands. These are offset 
from the dashed lines by an estimate of electromagnetic, $m_{u}\ne m_{d}$
and other systematic 
corrections to the masses. The experimental results are marked
at the physical $m_{d}/m_s$.}
\label{fig:masses}
\end{figure}

Our results are obtained from $u/d$ masses
larger than approximately three times the average~$m_{u/d}$ of the physical $u$~and $d$~quark masses. 
We obtain physical answers by extrapolating our results 
to the correct $m_{u/d}$ using chiral perturbation theory. 
In addition we have systematic errors from the 
finite lattice spacing values used. Since our 
results are so accurate we can also fit them as a function 
of $a$ to extrapolate to the physical $a=0$ limit. 
These two extrapolations are connected through the 
discretisation errors in the light quark action
and one way to treat those is by modifying chiral 
perturbation theory to handle them explicitly~\cite{fnal-fds}. 
A more general approach, that allows us to handle 
light and heavy quark discretisation errors together, 
is to perform a simultaneous fit for both 
chiral and continuum extrapolations allowing 
for expected functional forms 
in both with a 
Bayesian analysis~\cite{bayesfits} to constrain the coefficients. 
We tested this method by using it to analyze hundreds
of different fake datasets, generated using formulas 
from staggered chiral perturbation theory~\cite{schpt}
with random couplings. As expected, we found that 
roughly 70\% of the time the extrapolated results 
were within one standard deviation (computed using a 
Gaussian approximation to $\chi^2$) of the 
exact result from the formula, verifying the 
validity of our approach and of our error estimates. 

We fit our results to the standard continuum chiral expansions through first 
order~\cite{chiral-expansions}, augmented by second and third-order polynomial 
terms in $x_q\equiv B_0 m_q/8(\pi f_\pi)^2$, where $B_0 \equiv m_\pi^2/(m_u+m_d)$ 
to leading order in chiral perturbation theory. 
The polynomial corrections are 
required by the precision of our data~\footnote{ 
Our data are not sufficiently accurate to resolve logarithms from constants beyond first order.}. 
We include $D^*-D$ mass difference
terms in the $D/D_s$ chiral expansion and take the $DD^*\pi$ coupling 
to have the value inferred at leading order from the experimental $D^*$ width, allowing for 
a 30\% error from higher order effects. 
We correct for the finite volume of our 
lattice from chiral perturbation theory, 
although only $f_\pi$ has corrections larger than~0.5\%. 
Our corrections agree within 30\% 
with those in Ref.~\cite{colangelo} and we take a
50\% uncertainty in the correction. 
We fit the couplings in the chiral expansions simultaneously to our $\pi$ and $K$
masses and decay constants. We do the same for the masses 
and decay constants of the~$D$ and~$D_s$. 
Given the couplings, we tune $m_{u/d}$ and $m_s$ so that our formulas give 
the experimental values for $m_{\pi}$ and $m_K$ after correcting for 
the $u/d$ mass difference and electromagnetic effects
~\cite{milc3, dashen}.

We find that finite $a$ errors are 2--3.5~times smaller with 
the HISQ quark action than with the asqtad action, 
but still visible in our results. 
We combine the extrapolation to $a=0$ with the quark-mass extrapolation 
by adding $a^2$~dependence to our chiral formulas. 
We expect leading discretization errors of various types: $\alpha_s a^2$ and $a^4$ errors 
from conventional sources; 
and $\alpha_s^3 a^2$, $\alpha_s^3 a^2 \log(x_{u,d})$ and 
$\alpha_s^3 a^2 x_{u,d}$ from residual taste-changing interactions among the 
valence and sea light quarks. 
We do not have sufficient data to distinguish between these different functional 
forms, but we include all of them (with appropriate priors for their coefficients) in 
our fits so that uncertainties in the functional dependence on $a^2$ are correctly 
reflected in our final error analysis. The $a^2$~extrapolations are sufficiently small 
with HISQ (1\% or less for~$\pi$ and~$K$ from fine results to the 
continuum; 2\% for~$D$ and $D_s$) that 
the associated uncertainties in our final results are typically less than~0.5\%.
The combined chiral and continuum Bayesian fits have 45 parameters for 
$D/D_s$ and 48 for $\pi/K$ with 28 data points for each fit~\footnote{Reliable 
errors come from such fits by including realistic prior widths for all 
parameters. We take width 1 on $x_q^n$ and width $m_c^n$ on $a^n$ terms
in our chiral/continuum fits.}.

\begin{table}
\begin{ruledtabular}\begin{tabular}{lddddddd}
 & \multicolumn{1}{c}{$f_K/f_\pi$} & \multicolumn{1}{c}{$f_K$} & \multicolumn{1}{c}{$f_{\pi}$} & \multicolumn{1}{c}{$f_{D_s}/f_D$} & \multicolumn{1}{c}{$f_{D_s}$} & \multicolumn{1}{c}{$f_D$} & \multicolumn{1}{c}{$\Delta_s/\Delta_d$} \\ \hline
   $r_1$ uncerty. & 0.3 & 1.1 & 1.4 & 0.4 & 1.0 & 1.4 & 0.7 \\
    $a^2$ extrap. & 0.2 & 0.2 & 0.2 & 0.4 & 0.5 & 0.6 & 0.5 \\
       finite vol. & 0.4 & 0.4 & 0.8 & 0.3 & 0.1 & 0.3 & 0.1 \\
$m_{u/d}$ extrap. & 0.2 & 0.3 & 0.4 & 0.2 & 0.3 & 0.4 & 0.2 \\
  stat. errors & 0.2 & 0.4 & 0.5 & 0.5 & 0.6 & 0.7 & 0.6 \\
     $m_s$ evoln. & 0.1 & 0.1 & 0.1 & 0.3 & 0.3 & 0.3 & 0.5 \\
$m_d$, {\small QED} etc & 0.0 & 0.0 & 0.0 & 0.1 & 0.0 & 0.1 & 0.5 \\
\hline
               Total \% & 0.6 & 1.3 & 1.7 & 0.9 & 1.3 & 1.8 & 1.2 \\
\end{tabular}\end{ruledtabular}

\caption{Error budget (in \%) for our decay constants and 
mass ratio, where $\Delta_x = 2m_{D_x}-m_{\eta_c}$. 
The errors are defined so that it is 
easy to see how improvement will reduce them, e.g. the statistical uncertainty is the 
outcome of our fit, so that quadrupling statistics will halve it. 
The $a^2$ and $m_{u/d}$ 
extrapolation errors are the pieces of the Bayesian error that 
depend upon the prior widths in those extrapolations. 
`$m_s$ evolution' refers to the error 
in running the quark masses to the same scale from different $a$ 
values for the chiral extrapolation. 
The $r_1$ uncertainty comes from the error in 
the physical value of $r_1$ and the
finite volume uncertainty allows for a 50\% error in 
our finite volume adjustments described in the text. }
\label{tab:budget}
\end{table}

\begin{figure}[h]
\includegraphics{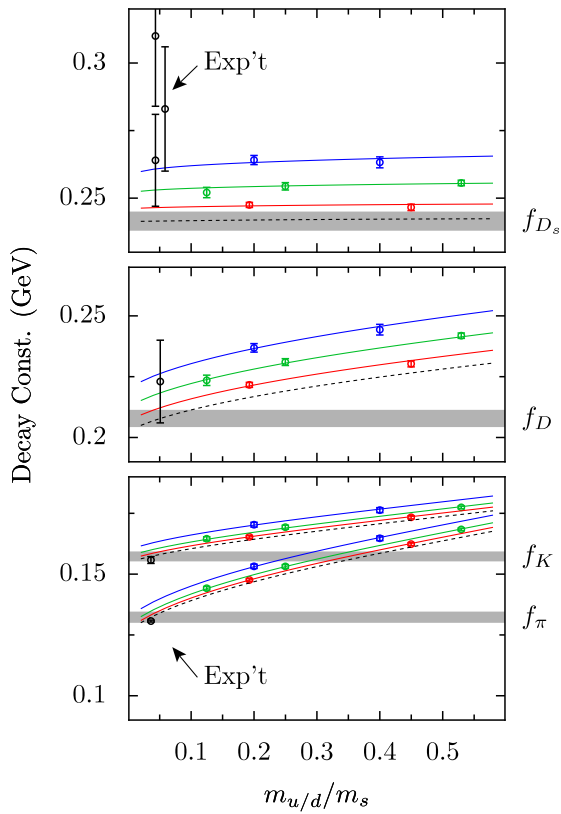}
\caption{$D$, $D_s$, $K$ and $\pi$ decay constants on 
very coarse, coarse and fine ensembles, as a function of the $u/d$ quark mass. 
The chiral fits are performed simultaneously with those 
of the corresponding meson masses, and the resulting continuum extrapolation 
curve is given by the dashed line. For $\pi, K$ we have $\chi^2/\mathrm{dof} = 0.2$ and 
for $D, D_s$, $\chi^2/\mathrm{dof}=0.6$, each for 28 pieces of data. The shaded band gives our final 
result. 
At the left are 
experimental 
results from CLEO-c~\cite{cleocfds, cleocfd} (with 
the $\tau$ decay result above the $\mu$ decay result for $D_s$) and BaBar~\cite{babar} ($D_s$ only)
and from the Particle Data Tables~\cite{pdg06} for $K$ and $\pi$. For the $K$ we have 
updated the result quoted by the PDG to be consistent with their quoted value of 
$V_{us}$. }
\label{fig:fds}
\end{figure}

Fig.~\ref{fig:masses} shows the masses of the 
$D$ and $D_s$ as a function of $u/d$ quark mass.
To reduce uncertainties from the scale and from $c$ quark mass tuning, 
the meson masses were obtained from $m_{D_q}-m_{\eta_c}/2+{m_{\eta_c}}_{expt}/2$.
The lines show our simultaneous chiral fits at each 
value of the lattice spacing and the dashed line the consequent extrapolation 
to $a=0$. The shaded bands give our final results: 
$m_{D_s}$ = 1.962(6) GeV, $m_D$ = 1.868(7) GeV. Experimental results 
are 1.968 GeV and 1.869 GeV respectively.  
We also obtain $(2m_{D_s}-m_{\eta_c})/(2m_D-m_{\eta_c})$ = 1.251(15), in excellent agreement with 
experiment, 1.260(2)~\cite{pdg06}. This last quantity is a non-trivial test of lattice QCD, since
we are accurately reproducing the difference in binding energies between 
a heavy-heavy state (the $\eta_c$ used to determine $m_c$) 
and a heavy-light state (the $D$ and $D_s$).  
Table~\ref{tab:budget} gives 
our complete error budget for this quantity. 

Fig.~\ref{fig:fds} similarly shows our results 
for decay constants on each ensemble with
complete error budgets 
in Table~\ref{tab:budget}.
$f_K$ and $f_{\pi}$ show very small discretisation 
effects and good agreement with experiment 
when $V_{ud}$ is taken from nuclear $\beta$ decay and 
$V_{us}$ from $K_{l3}$ decays~\cite{pdg06}. We obtain $f_{\pi}$ = 132(2) MeV and 
$f_K$ = 157(2) MeV. Alternatively
our result for $f_K/f_{\pi}$ (1.189(7)) can be used, with 
experimental leptonic branching fractions~\cite{marciano2, milc3}, to 
give $V_{us}$. Using the recent KLOE result for the $K$~\cite{kloe, blucher} we
obtain $V_{us}$ = 0.2262(13)(4) where the first error is theoretical and the 
second experimental. This agrees with, but improves on, the 
$K_{l3}$ result. Then $1-V_{ud}^2-V_{us}^2-V_{ub}^2$ = 0.0006(8), 
a precise test of CKM matrix first-row unitarity. 

$f_D$ and $f_{D_s}$ show larger discretisation 
effects but a more benign chiral extrapolation. 
Our final results are:
$f_{D_s}$ = 241(3) MeV, $f_D$ = 207(4) MeV and 
$f_{D_s}/f_D$ = 1.164(11). These results are 4--5 times more 
accurate than previous full lattice QCD results~\cite{fnal-fds} and existing 
experimental determinations. An interesting quantity is 
the double ratio $(f_{D_s}/f_D)/(f_K/f_{\pi})$. It is
estimated to be close to 1 from low
order chiral perturbation theory~\cite{becerevic}. 
We are able to make a strong quantitative statement 
with a value of 0.979(11). Equivalently $(\Phi_{D_s}/\Phi_D)/(f_K/f_{\pi})$ = 1.005(10), 
where $\Phi = f\sqrt{M}$. We also obtain 
$(f_{B_s}/f_B)/(f_{D_s}/f_D)$ = 1.03(3) using our previous 
result for the $B$ ratio~\cite{ourblept}. The B ratio dominates 
the error but improvement of this is underway.

The results for $f_D$ and $f_{D_s}$ obtained from the experimental 
leptonic branching rates coupled 
with CKM matrix elements determined from other processes 
(assuming $V_{cs} = V_{ud}$)
are also 
given in Fig.~\ref{fig:fds}. 
They are: $f_{D_s}$ = 264(17) MeV for $\mu$ decay 
and 310(26) MeV for $\tau$ decay from 
CLEO-c~\cite{cleocfds} and 283(23) MeV from BaBar~\cite{babar}, and $f_D$ =
223(17) MeV from CLEO-c for $\mu$ decay~\cite{cleocfd}. 
Using our results for $f_{D_s}$ and $f_{D_s}/f_D$ and the experimental 
values from CLEO-c~\cite{cleocfds} for $\mu$ decay (for 
consistency between $f_{D_s}$ and $f_D$)
we can directly
determine: $V_{cs} = 1.07(1)(7)$ and $V_{cs}/V_{cd}$ = 4.43(4)(41). 
The first error is theoretical and the second, and dominant one, experimental. 
The result for $V_{cs}$ improves on the direct 
determination of 0.96(9) given by the PDG~\cite{pdg06}. 

{\bf{Acknowledgements}} We are grateful to MILC for 
their configurations and to Quentin Mason and Doug Toussaint for 
useful discussions. The computing was done on Scotgrid and 
the QCDOCX cluster.
This work was supported by 
PPARC, the Royal Society, NSF and DoE. JS is grateful to Glasgow University and SUPA for hospitality while this work was completed.

\end{document}